\documentclass{nature}
\usepackage{amssymb}
\usepackage{latexsym}
\usepackage{graphicx}

\newcommand\apj{Astrophys. J.}

\newcommand\aap{Astron. Astrophys.}

\newcommand{\nicer}{\textit{NICER}}

\newcommand{\foru}{4U~1820-30}

\makeatletter
\let\saved@includegraphics\includegraphics
\AtBeginDocument{\let\includegraphics\saved@includegraphics}
\renewenvironment*{figure}{\@float{figure}}{\end@float}
\makeatother

\title{Interactions of Type I X-ray Bursts with Thin Accretion Discs}

\author{P. Chris Fragile$^{1,2}$, David R. Ballantyne$^3$ \& Aidan Blankenship$^1$}

\begin{document}

\maketitle

\begin{affiliations}
 \item Department of Physics \& Astronomy, College of Charleston, 66 George St., Charleston, SC 29424, USA
 \item Kavli Institute for Theoretical Physics, University of California Santa Barbara, Santa Barbara, CA 93106, USA
 \item Center for Relativistic Astrophysics, School of Physics, Georgia Institute of Technology, 837 State Street, Atlanta, GA 30332, USA
\end{affiliations}

\begin{abstract}
We perform a set of numerical experiments studying the interaction of Type I X-ray bursts with thin, Shakura-Sunyaev type accretion discs. Careful observations of X-ray spectra during such bursts have hinted at changes occurring in the inner regions of the disc. We now clearly demonstrate a number of key effects that take place simultaneously, including: evidence for weak, radiation-driven outflows along the surface of the disc; significant levels of Poynting-Robertson (PR) drag, leading to enhanced accretion; and prominent heating in the disc, which increases the height, while lowering the density and optical depth. The PR drag causes the inner edge of the disc to retreat from the neutron star surface toward larger radii and then recover on the timescale of the burst. We conclude that the rich interaction of an X-ray burst with the surrounding disc provides a novel way to study the physics of accretion onto compact objects.
\end{abstract}

Thermonuclear explosions on the surface of neutron stars, commonly known as Type I X-ray bursts, can be used to study the behavior of matter under extreme conditions\cite{Lewin93,Strohmayer06,Galloway18}. Careful analysis of the burst spectrum and luminosity may even provide constraints on the neutron star equation of state\cite{Lewin93,Guver12,Kajava14,Nattila17}, one of the most important unsolved problems in high-energy astrophysics. In addition, it has recently been recognized that X-ray bursts are a potentially powerful probe of accretion physics, as the intense release of radiative energy in the burst over a short timescale (seconds for Type I bursts; hours for superbursts) could significantly impact the structure of both the disc and corona\cite{Ballantyne05,intZand11,Degenaar18}. 

Observational evidence for the interaction of bursts with the accretion disc indicates a range of behaviors that signal a strong dependence on the accretion flow geometry, even for sources in the same spectral state. For example, several studies of bursts in the low/hard state\cite{Maccarone03,Chen13,Ji13} detected both a simultaneous rise in soft X-rays (owing to the burst itself) and drop in hard X-rays (attributed to cooling of a geometrically thick corona). On the other hand, bursts from other sources in the low/hard state produced reflection features in the X-ray spectra\cite{Ballantyne04,Keek14a,Keek14b,Keek18}, indicating the presence of an optically thick inner disc.

Recently, we presented the first numerical simulation of an accretion disc subject to the sudden, intense radiation field of an X-ray burst\cite{Fragile18b}. That simulation focused on a hot, geometrically-thick disc, and found that strong Compton cooling of the accreting plasma by the burst photons caused the disc temperature to drop by three orders of magnitude, the height to be reduced by one order of magnitude, and the accretion rate to increase by a factor of a few. All of these changes are consistent with many of the observed behaviors of burst-disc interactions in the low/hard state, though more work needs to be done to connect the simulations with observations. In the current work, we explore the complimentary case of thin accretion discs (corresponding to a more rapidly accreting system) interacting with an X-ray burst. 

\section{Results} 

\subsection{Retreat of $r_\mathrm{in}$.}
\begin{figure}
\begin{center}
\includegraphics[width=\textwidth]{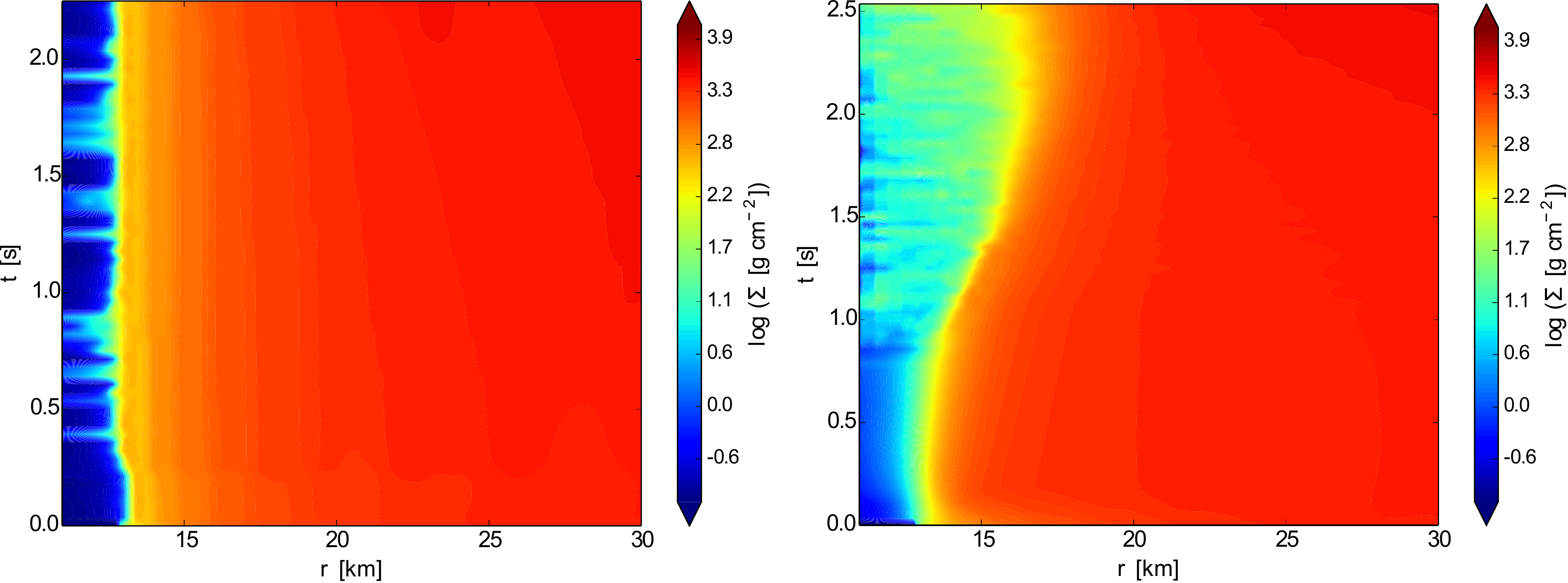}
\end{center}
\caption{{\bf Spacetime diagrams of the surface density.} {\it Left:} from the simulation without a burst; and {\it Right:} from the simulation with a burst of $3 \times 10^{38}$ erg~s$^{-1}$ showing a precipitous drop in the surface density on the timescale of the burst. Note that the simulation duration is different in the two panels.\label{fig:sigma}}
\end{figure}

\begin{figure}
\includegraphics[width=0.5\textwidth]{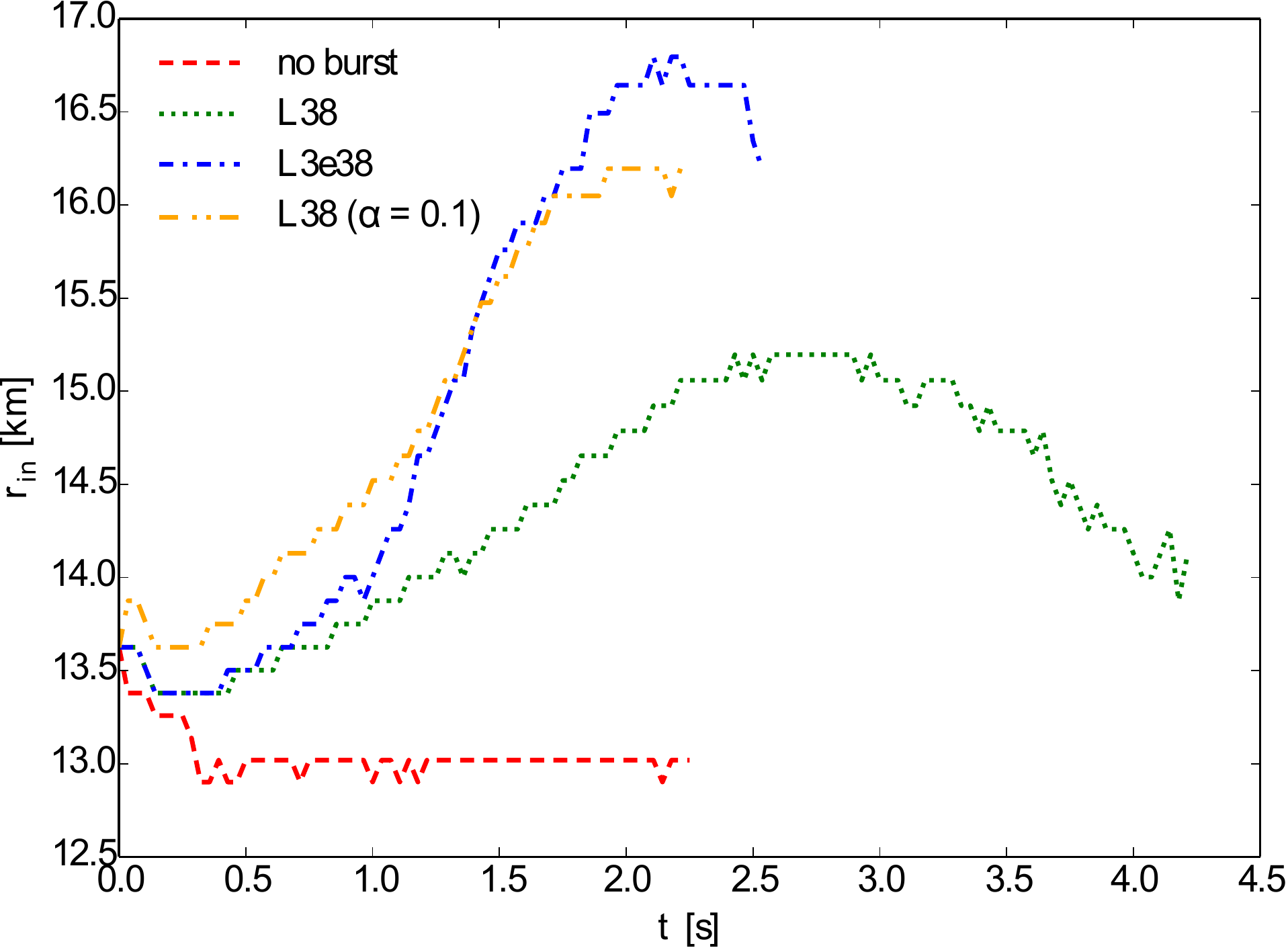}
\caption{{\bf Track of the inner radius of the accretion disc.} $r_\mathrm{in}$, defined as where $\Sigma(r) = \Sigma_0/3e$, as a function of time for all four simulations. While the no-burst simulation exhibits a nearly constant $r_\mathrm{in}$, the three burst simulations show a retreat of $r_\mathrm{in}$ on the timescale of the burst, followed by a recovery on a similar timescale.\label{fig:rin_sigma}}
\end{figure}

Based on previous observations\cite{Ballantyne04}, our expectation is that the inner edge of the disc, $r_\mathrm{in}$, will retreat (move outward) as a result of the burst. To assess this in our simulations, we begin with a measure of $r_\mathrm{in}$ that depends only on the hydrodynamic properties of the disc\cite{Fragile09}, specifically the surface density profile, $\Sigma(r)$. In Fig. \ref{fig:sigma}, we present spacetime diagrams of $\Sigma$ for simulations without and with a Type I X-ray burst. A noticeable retreat of the disc away from the neutron star is seen in all the simulations that include a burst, while no such movement is seen for the one simulation without. Formally defining the inner edge of the accretion disc, $r_\mathrm{in}$, as the radius where $\Sigma(r)$ drops from a reference value of $\Sigma_0 = 2.56\times 10^3\,\mathrm{g\,cm^{-2}}$ ($8.45\times 10^2\, \mathrm{g\,cm^{-2}}$ for the $\alpha = 0.1$ simulation), by a factor of $3e$, we can track the movement of $r_\mathrm{in}$ over the course of the burst, as we do in Fig. \ref{fig:rin_sigma}. While the simulation without a burst shows an initial slight decrease in $r_\mathrm{in}$, after which the value plateaus, all the simulations with bursts show $r_\mathrm{in}$ moving out by a few km and then back in over the timescale of the burst. We will return to the question of how quickly the disc fills back in and discuss mechanisms for this movement below. We also note that $\Delta r_\mathrm{in}$ is larger for our more luminous burst (blue, dot-dashed curve) and for the higher $\alpha$ simulation (golden, dot-dot-dashed curve).

Since these simulations include radiation, we can make a more relevant estimate of the inner radius of the disc by considering its effective optical depth, $\tau_\mathrm{e} = \sqrt{\kappa^\mathrm{a} \kappa^\mathrm{s}/2}\Sigma$ (see the Methods section for definitions of $\kappa^\mathrm{a}$ and $\kappa^\mathrm{s}$). Spacetime diagrams of $\tau_\mathrm{e}$ are shown in Fig. \ref{fig:tau_eff}. Now defining the inner edge of the accretion disc, $r_\mathrm{in}$, as the radius where $\tau_\mathrm{e}(r)$ drops below unity, we can again track the movement of $r_\mathrm{in}$ over the course of the burst. This track (not shown) agrees fairly well with the earlier results based on the disc surface density (Fig. \ref{fig:rin_sigma}). Again, the simulation without a burst shows only a very modest decrease in $r_\mathrm{in}$, whereas the simulations with bursts show the radius moving out by as much as 15-35\% near the peak of the burst, depending on its luminosity. Lastly, Fig.~\ref{fig:tau_eff} shows that the X-ray bursts reduce the optical depth over the entire inner accretion disc (not just the evacuated region) by an order of magnitude. This will have an important impact on how the burst-disc interaction is observed, a point we will return to below.

\subsection{Recovery of $r_\mathrm{in}$ after the burst.}
One thing we notice from Fig. \ref{fig:rin_sigma} is that the inner radius of the disc adjusts on roughly the same timescale as the burst itself. This is true both for the phase when the radius is moving outward (during the burst rise), but also during the phase when the disc recovers and moves inward (during the burst decay). This tells us that whatever mechanism causes $r_\mathrm{in}$ to change, it must act on a timescale shorter than the typical Type I burst timescales we are considering ($\sim$ few s). Certainly the dynamical timescale, $t_\mathrm{dyn}=2\pi/\Omega \approx 10^{-3}$ s, is much shorter than this over the regions of the disc being considered, as is the thermal timescale, $t_\mathrm{th} = t_\mathrm{dyn} / \alpha \approx 10^{-2}$ s, while the viscous timescale, $t_\mathrm{vis}=r^2/\nu = r^2/(\alpha H^2 \Omega) \approx 10^3$ s, is nominally much longer. This suggests that the disc does not have sufficient time to fill in viscously as the burst declines; it must be rebounding dynamically or thermally or be governed by some other timescale.

\subsection{Physical Cause of the Movement of $r_\mathrm{in}$.}
Earlier work\cite{Ballantyne05} had suggested three possible processes that could trigger the migration of $r_\mathrm{in}$ during a Type I X-ray burst: radiatively driven outflows, inflow due to Poynting-Robertson (PR) drag, and structural changes due to X-ray heating. We consider each of these in turn.

\subsection{Outflows.}
We first look at the mass flux through each radial shell in the simulation domain for simulations without and with bursts in the spacetime diagrams of Fig. \ref{fig:outflow}. While the simulation without a burst exhibits fairly constant mass inflow (light blue shades) at all radii over the duration of the simulation, the $L_0 = 10^{38}$ erg~s$^{-1}$ and $L_0 = 3 \times 10^{38}$ erg~s$^{-1}$ simulations transition from steady mass inflow to a mix of strong inflow (deep blue) and outflow (dark red), whereas the $L_0 = 10^{38}$ erg~s$^{-1}$, $\alpha = 0.1$ simulation mostly transitions to strong inflow. The strong inflows in the burst simulations are driven by a combination of PR drag and disc heating, which will be discussed in the following sections. The differences in the mass outflows comes from the convective stability of each simulation. While the no-burst and $L_0 = 10^{38}$ erg~s$^{-1}$, $\alpha = 0.1$ simulations remain stable under the Schwarzschild criterion,
\begin{equation}
\frac{dT}{dr} > \left(1 - \frac{1}{\Gamma}\right) \frac{T}{P}\frac{dP}{dr} ~,
\end{equation}
where $\Gamma=5/3$ is the polytropic index, the $L_0 = 10^{38}$ erg~s$^{-1}$ and $L_0 = 3 \times 10^{38}$ erg~s$^{-1}$ simulations do not.
 
 However, we can see in Fig. \ref{fig:rhoV} that much of the ``outflowing'' mass in the $L_0 = 10^{38}$ erg~s$^{-1}$ simulation (likewise for the $L_0 = 3 \times 10^{38}$ erg~s$^{-1}$ one) is located within the body of the disc. The only organized, true outflows we find in these simulations are confined to narrow layers along the top and bottom surfaces of the discs. These outflows are driven by the radiation pressure of the burst and achieve median speeds of $\approx 0.03c$. At these speeds, the matter is still formally bound to the neutron star and will fall back unless there is additional acceleration at larger radii. It is interesting that even in the case of the  $L_0 = 3 \times 10^{38}$ erg~s$^{-1}$ burst, which exceeds the Eddington limit of the accretion disc ($1.7 \times 10^{38}$~erg~s$^{-1}$), we do not find an appreciable outflow.

\subsection{Poynting-Robertson drag.}
 PR drag was seen to be significant in our previous thick-disc burst simulation\cite{Fragile18b} and appears to be present in the current thin-disc burst simulations, too. To illustrate this, Fig. \ref{fig:Vphi} plots the ratio of angular velocities for the simulation with the $3 \times 10^{38}$~erg~s$^{-1}$ burst, normalized by the one without. Like this case, all of the burst simulations show regions of sub-Keplerian flow near the inner edge of the disc around the time of the burst peak, $t_\mathrm{peak}$. This is reasonable, considering that the timescale for angular momentum loss due to PR drag is\cite{Walker89}
\begin{equation}
t_\mathrm{PR} \sim \frac{\Sigma c^2}{[\xi L/(4\pi r^2)]} ~,
\label{eq:PRtime}
\end{equation}
where $\xi$ is a geometric factor of order max[$dH/dr$, $R_\mathrm{NS}/r$]. Near the inner edges of the thin discs studied here, $\xi\sim 0.8$, and so this timescale is $t_\mathrm{PR} \sim 0.25$, $0.08$, and $0.08$ s for the $L_0 = 10^{38}$ erg s$^{-1}$, $L_0 = 3 \times 10^{38}$ erg~s$^{-1}$, and $L_0 = 10^{38}$ erg~s$^{-1}$ ($\alpha = 0.1$) simulations, respectively, which is about an order of magnitude shorter than the timescale of the burst. Thus, we expect PR drag to have sufficient time to act in the inner regions of the disc. In our previous thick disc simulations, the surface density was roughly six orders of magnitude smaller, corresponding to a similar reduction in the timescale for PR drag to act, explaining why this effect was noticeably stronger over a wider range of radii in those simulations.

The mass accretion rate associated with PR drag is\cite{Walker89}
\begin{equation}
\dot{M}_\mathrm{PR} \simeq 1.9\times 10^{17} \left(\frac{L_\mathrm{burst}}{L_\mathrm{Edd}}\right)~\mathrm{g~s}^{-1} ~.
\end{equation}
This is indeed very close to the peak mass accretion rate we measure (see Fig. 7). This mass accretion rate is not sustainable, though, as the inner part of the accretion disc is drained at a faster rate than it is being refilled, so $\dot{M}$ drops back down even before the burst peaks. Thus, the accretion rate onto the star is briefly enhanced by a factor of $\sim 10$, and has a sustained increase of a factor of a few. This behavior may explain similar increases in the accretion luminosities observed during X-ray bursts\cite{intZand13,Worpel13,Worpel15,Keek14a}. It is important to note that while PR drag can only affect the flow directly out to roughly $\tau_\mathrm{e} = 1$, the change in mass accretion rate at the inner edge of the disc can be effectively communicated throughout the rest of the disc by normal viscous torques.

\subsection{X-ray heating.}
The heating of the disc by the burst is significant for these thin discs. The radiation raises the density-weighted temperature by about half an order of magnitude over a wide range of radii (Fig. 8). This is interesting, because for thick discs, the radiation has the opposite effect, cooling the disc by orders of magnitude\cite{Fragile18b}. This makes sense, though, as the thin discs simulated here have non-burst, equilibrium temperatures that are cooler than the effective radiation temperature of the burst, whereas thick discs are initially much hotter\cite{Fragile18b}. In both cases, the disc is being driven toward thermodynamic equilibrium with the $\sim 10^7$ K burst radiation, mostly by Compton scattering.

Since our thin discs are gas-pressure supported, the heating also triggers a corresponding increase in the scale heights of the discs (Fig. 9). At any given radius, the height increases by almost a factor of two. According to standard thin disc theory, where
\begin{equation}
    \dot{M} \propto 3\pi\alpha H^2 \Omega\Sigma~,
\end{equation}
a doubling of $H$ (with $\Omega$ and $\Sigma$ remaining roughly constant) should lead to a fourfold increase in $\dot{M}$, which is, in fact, consistent with what we see at late times (Fig. 7). It is interesting to note that the temperature, height, and mass accretion rates have not recovered to their pre-burst values by the end of the simulations despite the simulations lasting many thermal timescales.

Because of the Kramer's style opacity we use, the rise in disc temperature and drop in density lead directly to a drop in the absorption opacity, which likewise leads to a drop in the effective optical depth, as seen in Fig. \ref{fig:tau_eff}. Here we are focused on the drop in optical depth {\em beyond} $r_\mathrm{in}$, i.e. in the body of the disc. The {\it movement} of $r_\mathrm{in}$ is owing to other effects, primarily PR drag.

\section{Discussion}
In this paper, we have considered the impact of Type-I X-ray bursts on surrounding thin accretion discs. Our simulations have produced a number of remarkable results:

(i) An order of magnitude drop in the effective optical depth of the disc.

(ii) Weak outflows along the surface of the disc.

(iii) Measurable Poynting-Robertson drag.

(iv) Half an order of magnitude increase in the average disc temperature.

(v) A factor of 2 increase in the disc height.

(vi) More than an order of magnitude increase in $\dot{M}$, initially driven by PR drag in the inner parts of the disc, tapering off to a factor of 4 enhancement over a broad range of radii due to the change in height of the disc.

Most remarkable, and consistent with observational evidence\cite{Ballantyne04,Keek14a,Keek14b}, we found that the inner radius of the disc moves away from the neutron star on the timescale of the burst. In our simulations, this was caused mostly by Poynting-Robertson-driven accretion, while other effects we considered, such as radiation-driven outflows, appeared to be negligible. Structural changes due to radiative heating of the disc were also observed, leading to the increase in height and decrease in optical depth mentioned above.

It is noteworthy that by all our measures, the inner edge of the accretion disc recovers (moves back in toward the neutron star) on roughly the same timescale as the decay of the burst. Since the viscous timescale is much longer than the burst timescale in these simulations, it would seem we can rule out the disc recovering via viscous replacement of lost material. Instead, the PR drag timescale, equation (\ref{eq:PRtime}), fits best with our results.

These are the first multi-dimensional, radiation hydrodynamic simulations of the interaction of Type I X-ray bursts with thin accretion discs, greatly improving on earlier one-dimensional treatments\cite{Walker92}. 
While there are still many improvements that can be made to the simulations, the physical mechanisms that lead to our main results appear robust. 

It may be possible in the future to combine such simulations with careful observations of bursts to constrain the properties of neutron star accretion discs, such as the viscosity parameter $\alpha$. Observationally, the interaction of an X-ray burst with a surrounding disc is most easily studied over the longer timescale of a superburst. However, our simulations are currently limited in duration to something more typical of a normal Type I burst ($\sim 10$ s).

Besides a longer burst, our results already show us that we can get a larger $\Delta r_\mathrm{in}$ whenever the luminosity is higher or the surface density is lower. This was the purpose of the $\alpha = 0.1$ simulation, since a larger $\alpha$ leads to a smaller $\Sigma$ for gas-pressure-dominated, Shakura-Sunyaev discs. The factor of 3 decrease in $\Sigma$ between the $L_0=10^{38}$ erg s$^{-1}$, $\alpha=0.025$ and $\alpha=0.1$ simulations produced a factor of 1.7 increase in $\Delta r_\mathrm{in}$. A smaller $\Sigma$ could also come from a lower mass accretion rate (assuming we are on the stable, gas-pressure-dominated branch of the Shakura-Sunyaev solution).

\begin{figure}
\begin{center}
\includegraphics[width=\textwidth]{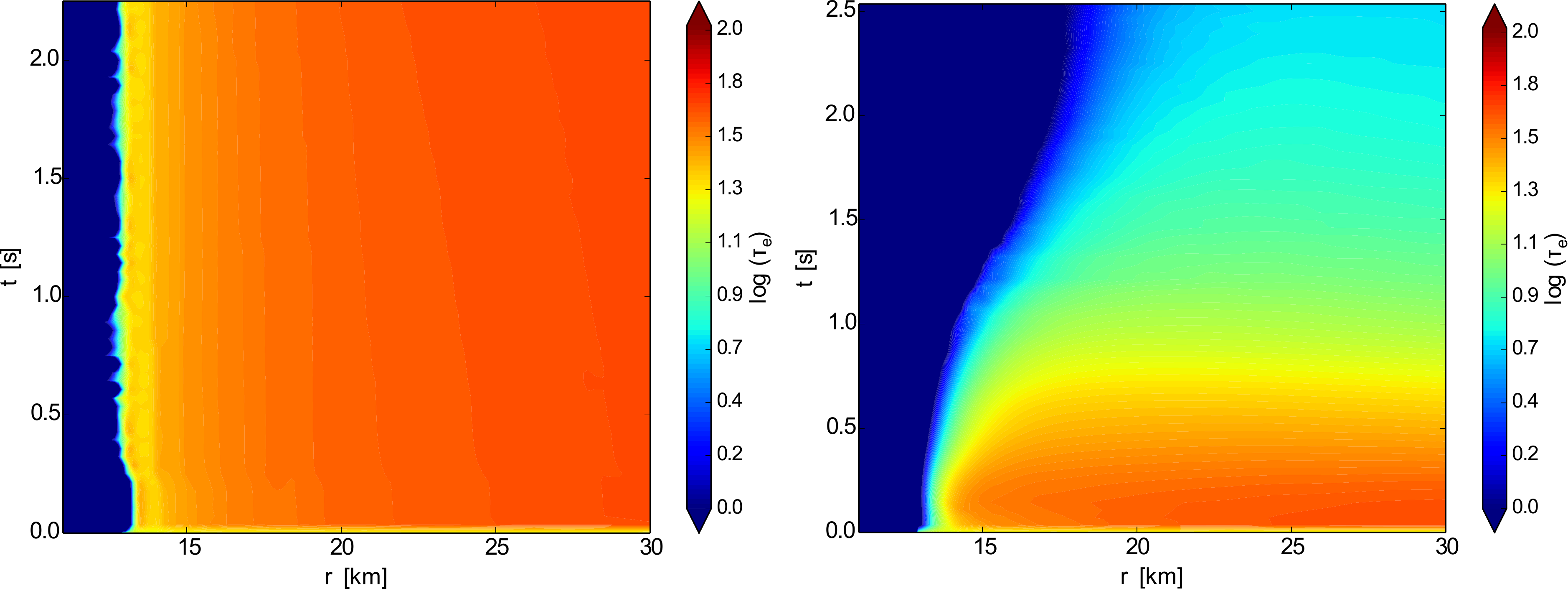}
\end{center}
\caption{{\bf Spacetime diagrams of the effective optical depth.} $\tau_\mathrm{e} (r,t) = \sqrt{\kappa^\mathrm{a} \kappa^\mathrm{s}/2}\Sigma$ {\it Left:} from the simulation without a burst; and {\it Right:} from the simulation with a burst of $3 \times 10^{38}$ erg~s$^{-1}$. The burst disc becomes optically thinner everywhere and transparent in its inner regions.\label{fig:tau_eff}}
\end{figure}

\begin{figure}
\begin{center}
\includegraphics[width=\textwidth]{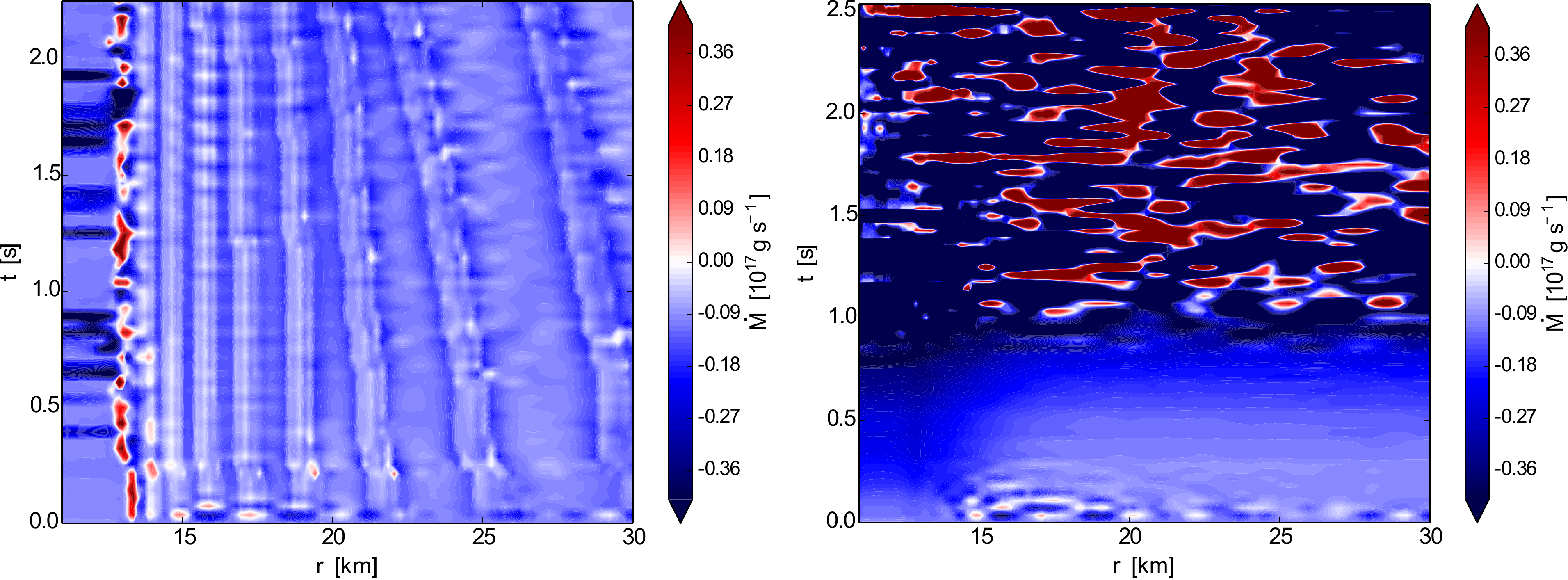}
\end{center}
\caption{{\bf Spacetime diagrams of the mass flux.} $\dot{M}$ through each radial shell {\it Left:} from the simulation without a burst; and {\it Right:} from the simulation with a burst of $3 \times 10^{38}$ erg~s$^{-1}$. While the no-burst simulation shows steady, modest accretion at all radii, the burst simulation exhibits rapid radial motion of gas both inward and outward.\label{fig:outflow}}
\end{figure}

\begin{figure}
\includegraphics[width=0.5\textwidth]{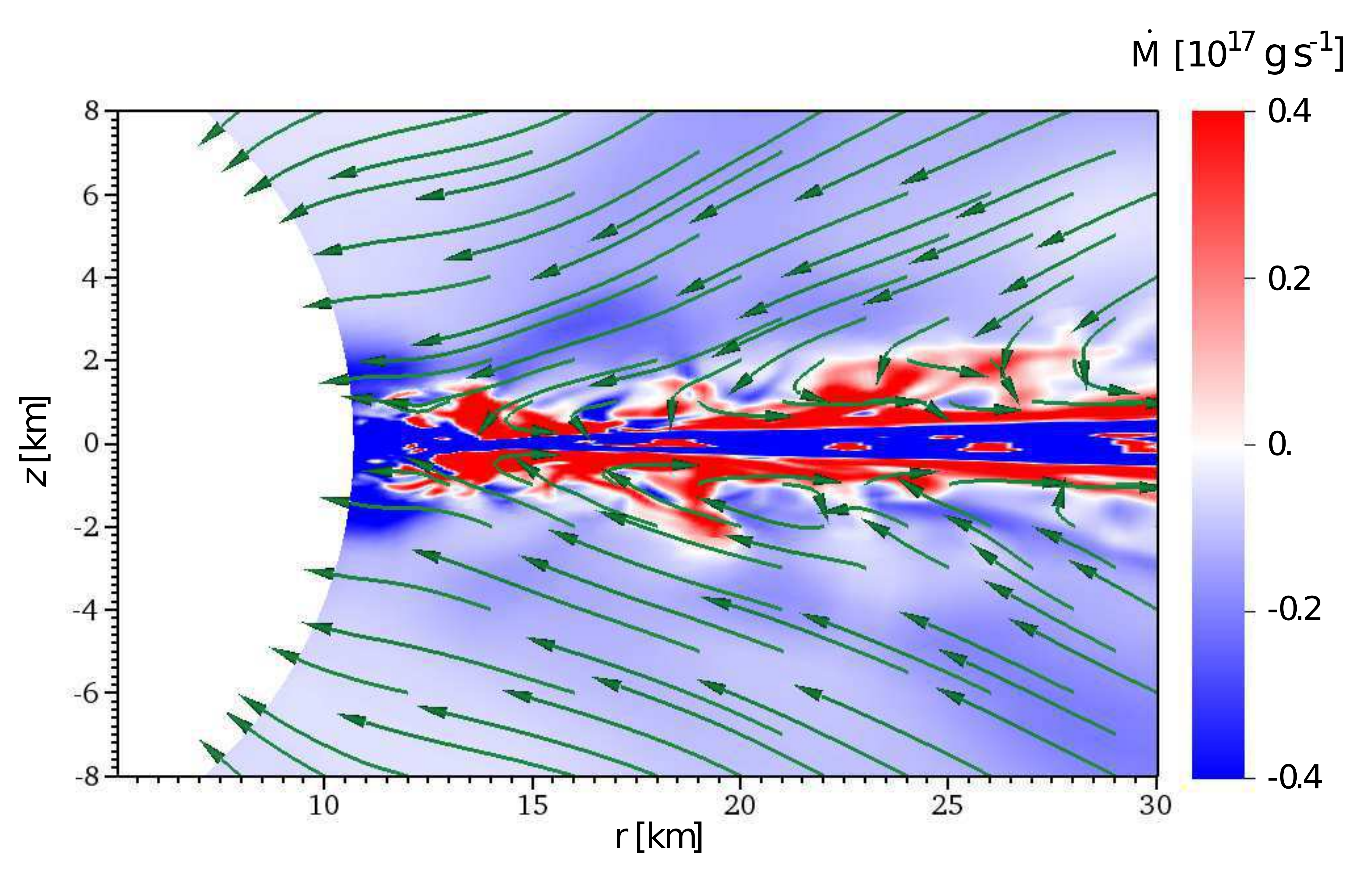}
\caption{{\bf Pseudocolor plot of the mass flux in the meridional plane with streamlines of the velocity field.} This plot is for the $10^{38}$ erg s$^{-1}$ ($\alpha = 0.025$) burst simulation, and data are averaged over the 0.5 s nearest the burst peak. Red colors correspond to outflowing material, while blue corresponds to inflowing. \label{fig:rhoV}}
\end{figure}

\begin{figure}
\includegraphics[width=0.5\textwidth]{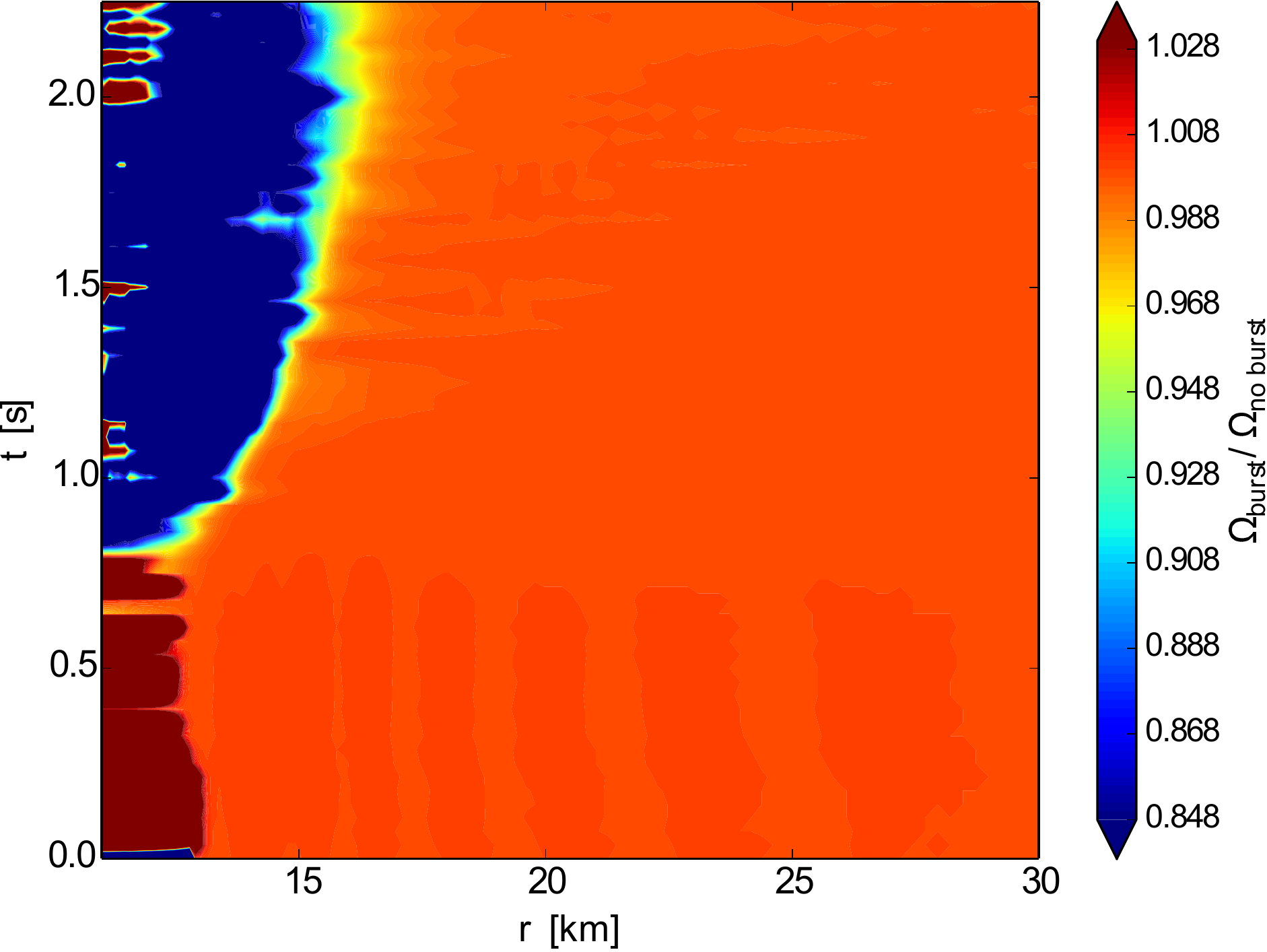}
\caption{{\bf Demonstration of Poynting-Robertson drag.} This spacetime diagram shows the ratio between the density-weighted orbital angular frequency for the $L_0 = 3 \times 10^{38}$ erg s$^{-1}$ simulation and the simulation without a Type I X-ray burst. Values of this ratio below 1 suggest some process is extracting angular momentum from the burst simulation.\label{fig:Vphi}}
\end{figure}

Future work will focus on the observational consequences of the predicted behavior. The illumination of the disc by the burst will cause X-ray reflection features\cite{Ball04} that can be used to track the changes to the disc geometry\cite{Ballantyne04,Keek14b}. Interestingly, the simulations show that X-ray heating causes an order of magnitude drop in the optical depth of the inner disc during the burst (Fig. \ref{fig:tau_eff}), which will reduce the reflection signal coming from the inner disc. As a result, despite the presence of the inner accretion disc, reflection features from the outer disc may dominate during a burst, as seen during the \foru\ superburst\cite{Ballantyne04}. Tracing the evolution of the disc reflection signal during an X-ray burst by a high-throughput spectral observatory such as \nicer\ or, in the future, \textit{eXTP}\cite{Zhang19} and \textit{STROBE-X}\cite{Ray19}, has the potential to resolve changes in the disc structure, elucidating key properties of accretion flows.



\bibliographystyle{naturemag}

\begin{thebibliography}{10}
\expandafter\ifx\csname url\endcsname\relax
  \def\url#1{\texttt{#1}}\fi
\expandafter\ifx\csname urlprefix\endcsname\relax\def\urlprefix{URL }\fi
\providecommand{\bibinfo}[2]{#2}
\providecommand{\eprint}[2][]{\url{#2}}

\bibitem{Lewin93}
\bibinfo{author}{{Lewin}, W.~H.~G.}, \bibinfo{author}{{van Paradijs}, J.} \&
  \bibinfo{author}{{Taam}, R.~E.}
\newblock \bibinfo{title}{{X-Ray Bursts}}.
\newblock \emph{\bibinfo{journal}{Sp. Sc. Rev.}} \textbf{\bibinfo{volume}{62}},
  \bibinfo{pages}{223--389} (\bibinfo{year}{1993}).

\bibitem{Strohmayer06}
\bibinfo{author}{{Strohmayer}, T.} \& \bibinfo{author}{{Bildsten}, L.}
\newblock \emph{\bibinfo{title}{{New views of thermonuclear bursts}}},
  \bibinfo{pages}{113--156} (\bibinfo{publisher}{Compact stellar X-ray
  sources}, \bibinfo{year}{2006}).

\bibitem{Galloway18}
\bibinfo{author}{{Galloway}, D.~K.} \& \bibinfo{author}{{Keek}, L.}
\newblock \bibinfo{title}{{Thermonuclear X-ray bursts}}
  \bibinfo{pages}{arXiv:1712.06227} (\bibinfo{year}{2017}).

\bibitem{Guver12}
\bibinfo{author}{{G{\"u}ver}, T.}, \bibinfo{author}{{Psaltis}, D.} \&
  \bibinfo{author}{{{\"O}zel}, F.}
\newblock \bibinfo{title}{{Systematic Uncertainties in the Spectroscopic
  Measurements of Neutron-star Masses and Radii from Thermonuclear X-Ray
  Bursts. I. Apparent Radii}}.
\newblock \emph{\bibinfo{journal}{Astrophys. J.}}
  \textbf{\bibinfo{volume}{747}}, \bibinfo{pages}{76} (\bibinfo{year}{2012}).

\bibitem{Kajava14}
\bibinfo{author}{{Kajava}, J.~J.~E.} \emph{et~al.}
\newblock \bibinfo{title}{{The influence of accretion geometry on the spectral
  evolution during thermonuclear (type I) X-ray bursts}}.
\newblock \emph{\bibinfo{journal}{Mon. Not. R. Astron. Soc.}}
  \textbf{\bibinfo{volume}{445}}, \bibinfo{pages}{4218--4234}
  (\bibinfo{year}{2014}).

\bibitem{Nattila17}
\bibinfo{author}{{N{\"a}ttil{\"a}}, J.} \emph{et~al.}
\newblock \bibinfo{title}{{Neutron star mass and radius measurements from
  atmospheric model fits to X-ray burst cooling tail spectra}}.
\newblock \emph{\bibinfo{journal}{\aap}} \textbf{\bibinfo{volume}{608}},
  \bibinfo{pages}{A31} (\bibinfo{year}{2017}).

\bibitem{Ballantyne05}
\bibinfo{author}{{Ballantyne}, D.~R.} \& \bibinfo{author}{{Everett}, J.~E.}
\newblock \bibinfo{title}{{On the Dynamics of Suddenly Heated Accretion Disks
  around Neutron Stars}}.
\newblock \emph{\bibinfo{journal}{Astrophys. J.}}
  \textbf{\bibinfo{volume}{626}}, \bibinfo{pages}{364--372}
  (\bibinfo{year}{2005}).

\bibitem{intZand11}
\bibinfo{author}{{in~'t~Zand}, J.~J.~M.}, \bibinfo{author}{{Galloway}, D.~K.}
  \& \bibinfo{author}{{Ballantyne}, D.~R.}
\newblock \bibinfo{title}{{Achromatic late-time variability in thermonuclear
  X-ray bursts. An accretion disk disrupted by a nova-like shell?}}
\newblock \emph{\bibinfo{journal}{\aap}} \textbf{\bibinfo{volume}{525}},
  \bibinfo{pages}{A111} (\bibinfo{year}{2011}).

\bibitem{Degenaar18}
\bibinfo{author}{{Degenaar}, N.} \emph{et~al.}
\newblock \bibinfo{title}{{Accretion Disks and Coronae in the X-Ray
  Flashlight}}.
\newblock \emph{\bibinfo{journal}{Sp. Sc. Rev.}}
  \textbf{\bibinfo{volume}{214}}, \bibinfo{pages}{15} (\bibinfo{year}{2018}).

\bibitem{Maccarone03}
\bibinfo{author}{{Maccarone}, T.~J.} \& \bibinfo{author}{{Coppi}, P.~S.}
\newblock \bibinfo{title}{{Spectral fits to the 1999 Aql X-1 outburst data}}.
\newblock \emph{\bibinfo{journal}{\aap}} \textbf{\bibinfo{volume}{399}},
  \bibinfo{pages}{1151--1157} (\bibinfo{year}{2003}).

\bibitem{Chen13}
\bibinfo{author}{{Chen}, Y.-P.} \emph{et~al.}
\newblock \bibinfo{title}{{The Hard X-Ray Behavior of Aql X-1 during Type-I
  Bursts}}.
\newblock \emph{\bibinfo{journal}{Astrophys. J.}}
  \textbf{\bibinfo{volume}{777}}, \bibinfo{pages}{L9} (\bibinfo{year}{2013}).

\bibitem{Ji13}
\bibinfo{author}{{Ji}, L.} \emph{et~al.}
\newblock \bibinfo{title}{{X-ray bursts as a probe of the corona: the case of
  XRB 4U 1636-536}}.
\newblock \emph{\bibinfo{journal}{Mon. Not. R. Astron. Soc.}}
  \textbf{\bibinfo{volume}{432}}, \bibinfo{pages}{2773--2778}
  (\bibinfo{year}{2013}).

\bibitem{Ballantyne04}
\bibinfo{author}{{Ballantyne}, D.~R.} \& \bibinfo{author}{{Strohmayer}, T.~E.}
\newblock \bibinfo{title}{{The Evolution of the Accretion Disk around 4U
  1820-30 during a Superburst}}.
\newblock \emph{\bibinfo{journal}{Astrophys. J.}}
  \textbf{\bibinfo{volume}{602}}, \bibinfo{pages}{L105--L108}
  (\bibinfo{year}{2004}).

\bibitem{Keek14a}
\bibinfo{author}{{Keek}, L.}, \bibinfo{author}{{Ballantyne}, D.~R.},
  \bibinfo{author}{{Kuulkers}, E.} \& \bibinfo{author}{{Strohmayer}, T.~E.}
\newblock \bibinfo{title}{{Characterizing the Evolving X-Ray Spectral Features
  during a Superburst from 4U 1636-536}}.
\newblock \emph{\bibinfo{journal}{Astrophys. J.}}
  \textbf{\bibinfo{volume}{789}}, \bibinfo{pages}{121} (\bibinfo{year}{2014}).

\bibitem{Keek14b}
\bibinfo{author}{{Keek}, L.}, \bibinfo{author}{{Ballantyne}, D.~R.},
  \bibinfo{author}{{Kuulkers}, E.} \& \bibinfo{author}{{Strohmayer}, T.~E.}
\newblock \bibinfo{title}{{X-Raying an Accretion Disk in Realtime: The
  Evolution of Ionized Reflection during a Superburst from 4U 1636-536}}.
\newblock \emph{\bibinfo{journal}{Astrophys. J.}}
  \textbf{\bibinfo{volume}{797}}, \bibinfo{pages}{L23} (\bibinfo{year}{2014}).

\bibitem{Keek18}
\bibinfo{author}{{Keek}, L.} \emph{et~al.}
\newblock \bibinfo{title}{{NICER Observes the Effects of an X-Ray Burst on the
  Accretion Environment in Aql X-1}}.
\newblock \emph{\bibinfo{journal}{Astrophys. J.}}
  \textbf{\bibinfo{volume}{855}}, \bibinfo{pages}{L4} (\bibinfo{year}{2018}).

\bibitem{Fragile18b}
\bibinfo{author}{{Fragile}, P.~C.}, \bibinfo{author}{{Ballantyne}, D.~R.},
  \bibinfo{author}{{Maccarone}, T.~J.} \& \bibinfo{author}{{Witry}, J.~W.~L.}
\newblock \bibinfo{title}{{Simulating the Collapse of a Thick Accretion Disk
  due to a Type I X-Ray Burst from a Neutron Star}}.
\newblock \emph{\bibinfo{journal}{Astrophys. J.}}
  \textbf{\bibinfo{volume}{867}}, \bibinfo{pages}{L28} (\bibinfo{year}{2018}).

\bibitem{Fragile09}
\bibinfo{author}{{Fragile}, P.~C.}
\newblock \bibinfo{title}{{Effective Inner Radius of Tilted Black Hole
  Accretion Disks}}.
\newblock \emph{\bibinfo{journal}{Astrophys. J.}}
  \textbf{\bibinfo{volume}{706}}, \bibinfo{pages}{L246--L250}
  (\bibinfo{year}{2009}).

\bibitem{Walker89}
\bibinfo{author}{{Walker}, M.~A.} \& \bibinfo{author}{{Meszaros}, P.}
\newblock \bibinfo{title}{{The dynamical influence of radiation in type 1 X-ray
  bursts}}.
\newblock \emph{\bibinfo{journal}{Astrophys. J.}}
  \textbf{\bibinfo{volume}{346}}, \bibinfo{pages}{844--846}
  (\bibinfo{year}{1989}).

\bibitem{intZand13}
\bibinfo{author}{{in't Zand}, J.~J.~M.} \emph{et~al.}
\newblock \bibinfo{title}{{A bright thermonuclear X-ray burst simultaneously
  observed with Chandra and RXTE}}.
\newblock \emph{\bibinfo{journal}{\aap}} \textbf{\bibinfo{volume}{553}},
  \bibinfo{pages}{A83} (\bibinfo{year}{2013}).

\bibitem{Worpel13}
\bibinfo{author}{{Worpel}, H.}, \bibinfo{author}{{Galloway}, D.~K.} \&
  \bibinfo{author}{{Price}, D.~J.}
\newblock \bibinfo{title}{{Evidence for Accretion Rate Change during Type I
  X-Ray Bursts}}.
\newblock \emph{\bibinfo{journal}{Astrophys. J.}}
  \textbf{\bibinfo{volume}{772}}, \bibinfo{pages}{94} (\bibinfo{year}{2013}).

\bibitem{Worpel15}
\bibinfo{author}{{Worpel}, H.}, \bibinfo{author}{{Galloway}, D.~K.} \&
  \bibinfo{author}{{Price}, D.~J.}
\newblock \bibinfo{title}{{Evidence for Enhanced Persistent Emission During
  Sub-Eddington Thermonuclear Bursts}}.
\newblock \emph{\bibinfo{journal}{Astrophys. J.}}
  \textbf{\bibinfo{volume}{801}}, \bibinfo{pages}{60} (\bibinfo{year}{2015}).

\bibitem{Walker92}
\bibinfo{author}{{Walker}, M.~A.}
\newblock \bibinfo{title}{{Radiation Dynamics in X-Ray Binaries. I. Type 1
  Bursts}}.
\newblock \emph{\bibinfo{journal}{\apj}} \textbf{\bibinfo{volume}{385}},
  \bibinfo{pages}{642} (\bibinfo{year}{1992}).

\bibitem{Ball04}
\bibinfo{author}{{Ballantyne}, D.~R.}
\newblock \bibinfo{title}{{Reflection spectra from an accretion disc
  illuminated by a neutron star X-ray burst}}.
\newblock \emph{\bibinfo{journal}{Mon. Not. R. Astron. Soc.}}
  \textbf{\bibinfo{volume}{351}}, \bibinfo{pages}{57--62}
  (\bibinfo{year}{2004}).

\bibitem{Zhang19}
\bibinfo{author}{{Zhang}, S.} \emph{et~al.}
\newblock \bibinfo{title}{{The enhanced X-ray Timing and Polarimetry
  mission{\textemdash}eXTP}}.
\newblock \emph{\bibinfo{journal}{Science China Physics, Mechanics, and
  Astronomy}} \textbf{\bibinfo{volume}{62}}, \bibinfo{pages}{29502}
  (\bibinfo{year}{2019}).

\bibitem{Ray19}
\bibinfo{author}{{Ray}, P.~S.} \emph{et~al.}
\newblock \bibinfo{title}{{STROBE-X: X-ray Timing and Spectroscopy on Dynamical
  Timescales from Microseconds to Years}} \bibinfo{pages}{arXiv:1903.03035}
  (\bibinfo{year}{2019}).

\bibitem{Anninos05}
\bibinfo{author}{{Anninos}, P.}, \bibinfo{author}{{Fragile}, P.~C.} \&
  \bibinfo{author}{{Salmonson}, J.~D.}
\newblock \bibinfo{title}{{Cosmos++: Relativistic Magnetohydrodynamics on
  Unstructured Grids with Local Adaptive Refinement}}.
\newblock \emph{\bibinfo{journal}{Astrophys. J.}}
  \textbf{\bibinfo{volume}{635}}, \bibinfo{pages}{723--740}
  (\bibinfo{year}{2005}).

\bibitem{Fragile12}
\bibinfo{author}{{Fragile}, P.~C.}, \bibinfo{author}{{Gillespie}, A.},
  \bibinfo{author}{{Monahan}, T.}, \bibinfo{author}{{Rodriguez}, M.} \&
  \bibinfo{author}{{Anninos}, P.}
\newblock \bibinfo{title}{{Numerical Simulations of Optically Thick Accretion
  onto a Black Hole. I. Spherical Case}}.
\newblock \emph{\bibinfo{journal}{Astrophys. J. Suppl. Ser.}}
  \textbf{\bibinfo{volume}{201}}, \bibinfo{pages}{9} (\bibinfo{year}{2012}).

\bibitem{Fragile14}
\bibinfo{author}{{Fragile}, P.~C.}, \bibinfo{author}{{Olejar}, A.} \&
  \bibinfo{author}{{Anninos}, P.}
\newblock \bibinfo{title}{{Numerical Simulations of Optically Thick Accretion
  onto a Black Hole. II. Rotating Flow}}.
\newblock \emph{\bibinfo{journal}{Astrophys. J.}}
  \textbf{\bibinfo{volume}{796}}, \bibinfo{pages}{22} (\bibinfo{year}{2014}).

\bibitem{Fragile18a}
\bibinfo{author}{{Fragile}, P.~C.}, \bibinfo{author}{{Etheridge}, S.~M.},
  \bibinfo{author}{{Anninos}, P.}, \bibinfo{author}{{Mishra}, B.} \&
  \bibinfo{author}{{Klu{\'z}niak}, W.}
\newblock \bibinfo{title}{{Relativistic, Viscous, Radiation Hydrodynamic
  Simulations of Geometrically Thin Disks. I. Thermal and Other
  Instabilities}}.
\newblock \emph{\bibinfo{journal}{Astrophys. J.}}
  \textbf{\bibinfo{volume}{857}}, \bibinfo{pages}{1} (\bibinfo{year}{2018}).

\bibitem{Miller15}
\bibinfo{author}{{Miller}, M.~C.} \& \bibinfo{author}{{Miller}, J.~M.}
\newblock \bibinfo{title}{{The masses and spins of neutron stars and
  stellar-mass black holes}}.
\newblock \emph{\bibinfo{journal}{Phys. Rep}} \textbf{\bibinfo{volume}{548}},
  \bibinfo{pages}{1--34} (\bibinfo{year}{2015}).

\bibitem{Levermore84}
\bibinfo{author}{{Levermore}, C.~D.}
\newblock \bibinfo{title}{{Relating Eddington factors to flux limiters.}}
\newblock \emph{\bibinfo{journal}{J. Quant. Spectrosc. Radiat. Transf.}}
  \textbf{\bibinfo{volume}{31}}, \bibinfo{pages}{149--160}
  (\bibinfo{year}{1984}).

\bibitem{Sadowski13}
\bibinfo{author}{{S{\c a}dowski}, A.}, \bibinfo{author}{{Narayan}, R.},
  \bibinfo{author}{{Tchekhovskoy}, A.} \& \bibinfo{author}{{Zhu}, Y.}
\newblock \bibinfo{title}{{Semi-implicit scheme for treating radiation under M1
  closure in general relativistic conservative fluid dynamics codes}}.
\newblock \emph{\bibinfo{journal}{Mon. Not. R. Astron. Soc.}}
  \textbf{\bibinfo{volume}{429}}, \bibinfo{pages}{3533--3550}
  (\bibinfo{year}{2013}).

\bibitem{Hirose09}
\bibinfo{author}{{Hirose}, S.}, \bibinfo{author}{{Krolik}, J.~H.} \&
  \bibinfo{author}{{Blaes}, O.}
\newblock \bibinfo{title}{{Radiation-Dominated Disks are Thermally Stable}}.
\newblock \emph{\bibinfo{journal}{Astrophys. J.}}
  \textbf{\bibinfo{volume}{691}}, \bibinfo{pages}{16--31}
  (\bibinfo{year}{2009}).

\bibitem{Narayan95}
\bibinfo{author}{{Narayan}, R.} \& \bibinfo{author}{{Yi}, I.}
\newblock \bibinfo{title}{{Advection-dominated Accretion: Underfed Black Holes
  and Neutron Stars}}.
\newblock \emph{\bibinfo{journal}{Astrophys. J.}}
  \textbf{\bibinfo{volume}{452}}, \bibinfo{pages}{710} (\bibinfo{year}{1995}).

\bibitem{Norris05}
\bibinfo{author}{{Norris}, J.~P.} \emph{et~al.}
\newblock \bibinfo{title}{{Long-Lag, Wide-Pulse Gamma-Ray Bursts}}.
\newblock \emph{\bibinfo{journal}{Astrophys. J.}}
  \textbf{\bibinfo{volume}{627}}, \bibinfo{pages}{324--345}
  (\bibinfo{year}{2005}).

\bibitem{Novikov73}
\bibinfo{author}{{Novikov}, I.~D.} \& \bibinfo{author}{{Thorne}, K.~S.}
\newblock \bibinfo{title}{{Astrophysics of black holes.}}
\newblock In \bibinfo{editor}{{Dewitt}, C.} \& \bibinfo{editor}{{Dewitt},
  B.~S.} (eds.) \emph{\bibinfo{booktitle}{Black Holes (Les Astres Occlus)}},
  \bibinfo{pages}{343--450} (\bibinfo{year}{1973}).

\bibitem{Abramowicz13}
\bibinfo{author}{{Abramowicz}, M.~A.} \& \bibinfo{author}{{Fragile}, P.~C.}
\newblock \bibinfo{title}{{Foundations of Black Hole Accretion Disk Theory}}.
\newblock \emph{\bibinfo{journal}{Living Rev. Rel.}}
  \textbf{\bibinfo{volume}{16}}, \bibinfo{pages}{1} (\bibinfo{year}{2013}).

\bibitem{Penna12}
\bibinfo{author}{{Penna}, R.~F.}, \bibinfo{author}{{S{\c a}owski}, A.} \&
  \bibinfo{author}{{McKinney}, J.~C.}
\newblock \bibinfo{title}{{Thin-disc theory with a non-zero-torque boundary
  condition and comparisons with simulations}}.
\newblock \emph{\bibinfo{journal}{Mon. Not. R. Astron. Soc.}}
  \textbf{\bibinfo{volume}{420}}, \bibinfo{pages}{684--698}
  (\bibinfo{year}{2012}).

\bibitem{Chandrasekhar60}
\bibinfo{author}{{Chandrasekhar}, S.}
\newblock \emph{\bibinfo{title}{{Radiative transfer}}} (\bibinfo{publisher}{New
  York: Dover}, \bibinfo{year}{1960}).

\bibitem{Lapidus85}
\bibinfo{author}{{Lapidus}, I.~I.} \& \bibinfo{author}{{Sunyaev}, R.~A.}
\newblock \bibinfo{title}{{Angular distribution and polarization of
  X-ray-burster radiation (during stationary and flash phases)}}.
\newblock \emph{\bibinfo{journal}{Mon. Not. R. Astron. Soc.}}
  \textbf{\bibinfo{volume}{217}}, \bibinfo{pages}{291--303}
  (\bibinfo{year}{1985}).

\bibitem{Mahmoodifar16}
\bibinfo{author}{{Mahmoodifar}, S.} \& \bibinfo{author}{{Strohmayer}, T.}
\newblock \bibinfo{title}{{X-Ray Burst Oscillations: From Flame Spreading to
  the Cooling Wake}}.
\newblock \emph{\bibinfo{journal}{Astrophys. J.}}
  \textbf{\bibinfo{volume}{818}}, \bibinfo{pages}{93} (\bibinfo{year}{2016}).

\end{thebibliography}


\begin{addendum}
 \item P.C.F. and A.B. acknowledge support from SC NASA EPSCoR RGP 2017 and National Science Foundation grants AST-1616185 and AST-1907850. P.C.F. acknowledges support from National Science Foundation grant PHY-1748958. A.B. acknowledges support from the College of Charleston Undergraduate Research and Creative Activities Board, through SURF grant SU2019-01. This work used the Extreme Science and Engineering Discovery Environment (XSEDE), which is supported by National Science Foundation grant number ACI-1548562.
 \item[Author contributions] P.C.F. wrote the manuscript with input from all authors. P.C.F. and D.R.B. wrote the funding proposal that supported this work. P.C.F. and A.B. designed and executed the simulations and analyzed the results.
 \item[Competing interests] The authors declare that they have no competing interests.
 \item[Correspondence and requests for materials] should be addressed to P.C.F.~(email: fragilep@cofc.edu).
\end{addendum}

\begin{methods}
The simulations use the {\em Cosmos++} computational astrophysics code\cite{Anninos05,Fragile12,Fragile14} to numerically evolve the equations of general relativistic radiative, viscous hydrodynamics\cite{Fragile18a}. We use the Kerr metric, with mass $M = 1.45 M_{\odot}$ and spin parameter $a = 0$, to describe the spacetime exterior to the neutron star. This is something that should be explored further in future work, as PR drag is known to be a significant function of spin\cite{Walker92} and accreting neutron stars may have dimensionless spins as high as 0.3 \cite{Miller15}.

For the radiation, we employ a covariant formulation of the $\bf{M}_1$ closure scheme\cite{Levermore84,Sadowski13,Fragile14}. In this work, we assume Kramers-type, gray opacity laws. Thus, the burst spectral energy distribution is not taken into account. Since free-free absorption is the most relevant atomic absorption process, the appropriate Planck and Rosseland means (for solar metallicity and a hydrogen mass fraction of $X=0.7$) are $\kappa_\mathrm{P}^\mathrm{a} = 6.4\times 10^{22} T^{-7/2}_\mathrm{K} \rho_\mathrm{cgs} ~\mathrm{cm}^2~\mathrm{g}^{-1}$ and $\kappa_\mathrm{R}^\mathrm{a} = 1.6\times 10^{21} T^{-7/2}_\mathrm{K} \rho_\mathrm{cgs} ~\mathrm{cm}^2~\mathrm{g}^{-1}$ \cite{Hirose09}, respectively, where $T_\mathrm{K}$ is the ideal gas temperature of the fluid in Kelvin and $\rho_\mathrm{cgs}$ is the density in g cm$^{-3}$. In this work, we assume the flux mean, $\kappa_\mathrm{F}^\mathrm{a}$, is the same as the Rosseland mean and the J-mean, $\kappa_\mathrm{J}^\mathrm{a}$, is the same as the Planck mean. We include a Compton-scattering correction, $4 \kappa^\mathrm{s}(T_\mathrm{gas} - T_\mathrm{rad})/m_e$, and use a constant electron scattering opacity, $\kappa^\mathrm{s} = 0.2(1+X) = 0.34 ~\mathrm{cm}^2~\mathrm{g}^{-1}$.  Further, we assume that the electron-ion equilibration time is sufficiently short for the electrons to be at the same temperature as the ions.

Following standard disc theory, the shear viscosity coefficient is calculated as 
\begin{equation}
\mu=\nu\rho=\alpha\rho c_s H ~,
\end{equation}
where $c_s$ is the thermal sound speed (including both gas and radiation contributions), $H$ is the disc height, and $\alpha$ is the Shakura-Sunyaev viscosity parameter.  In this work, $\alpha$ is assumed to be a constant, while $c_s$ and $H$ are evaluated from local conditions within the fluid, except that $H$ is limited to $< 0.1 r$ to prevent having very high viscosity in the background gas where $V^\phi$ fluctuates considerably. For most simulations, we choose $\alpha = 0.025$, though for one simulation we consider $\alpha = 0.1$ in order to study the effects of the larger viscosity.

Through viscous heating and radiative cooling, the thermodynamics and radiative luminosity of the disc are accounted for self-consistently within the simulations. We further include radiation coming from the surface of the neutron star by including a variable flux at the inner radial boundary of our simulation domain. This flux is composed of radiation owing to accretion of matter onto the surface of the neutron star and (when appropriate) to radiation from a Type I X-ray burst on the surface. For the accretion luminosity, we assume $L_\mathrm{surf} = 0.14 \dot{M} c^2$ (corresponding to the additional potential energy lost as material falls from the inner edge of the disc to the surface of the neutron star\cite{Narayan95}), where
\begin{equation}
    \dot{M} = 2\pi \int_0^\pi \sqrt{-g} \rho u^r d\theta ~,
\end{equation}
$g$ is the determinant of the curvature metric, and $u^r$ is the radial component of the fluid 4-velocity. 

Since most Type I X-ray bursts exhibit a fast rise followed by a slower decay, similar in shape to many gamma-ray burst (GRB) pulse profiles, we model the burst luminosity, $L_\mathrm{burst}$, using the so-called Norris model from the GRB community\cite{Norris05}:
\begin{equation}
L_\mathrm{burst}(t) = L_0 e^{2(\tau_1/\tau_2)^{1/2}} e^{\frac{-\tau_1}{t-t_s} - \frac{t-t_s}{\tau_2}} ~,
\label{eqn:Norris}
\end{equation}
where $L_0$ is the peak burst luminosity, $t_s$ is the burst start time, and $\tau_1$ and $\tau_2$ characterize the burst rise and decay. In this work we choose $t_s = -0.4$~s, $\tau_1 = 6$~s, and $\tau_2 = 1$~s and consider two values of $L_0$ ($10^{38}$ and $3\times 10^{38}$~erg~s$^{-1}$ or $0.59 L_\mathrm{Edd}$ and $1.76 L_\mathrm{Edd}$, respectively, where $L_\mathrm{Edd} = 1.7 \times 10^{38}$ erg~s$^{-1}$). These choices produce bursts that peak at a simulation time of $t=2.05$s, last for 10s, and have total energy outputs of $3$ and $9\times 10^{38}$~erg, respectively. Profiles for each of our bursts are illustrated in Fig. 10. Each of our simulations follows $>2$~s of the burst, ensuring that the peak of the lightcurve is captured by the calculations. In addition, the $10^{38}$~erg~s$^{-1}$, $\alpha=0.025$ burst is followed for $>4$~s in order to measure changes in the disc during the burst tail. Note that even though we consider one case of a super-Eddington burst, we do not include the effects of the expansion of the neutron star photosphere that is expected to accompany this case. This will be a priority in future simulations.
 
To initialize the discs, we start from the relativistic generalization of the Shakura-Sunyaev thin disc solution\cite{Novikov73,Abramowicz13}. As we are only considering a limited radial range, and to avoid any thermal or other instabilities\cite{Fragile18a}, we only consider the gas-pressure-dominated regime (appropriate for $\dot{M}c^2/L_\mathrm{Edd} \lesssim 0.02$). We also include a small radial drift velocity, $V^r(r)$ \cite{Penna12}.

For the initial vertical profile, we solve for the vertical hydrostatic equilibrium, assuming an isothermal disc:
\begin{equation}
\rho(r,z) = \rho_0 \mathrm{e}^{-z^2/2H^2} 
\end{equation}
and
\begin{equation}
P_\mathrm{tot}(r,z) = \frac{G M H^2}{r^3}\rho(r,z)~.
\end{equation}
Assuming the gas and radiation are in local thermodynamic equilibrium for the initial, analytic solution, we partition the pressure according to
\begin{equation}
P_\mathrm{tot} = P_\mathrm{gas} + P_\mathrm{rad} = \frac{k_\mathrm{b}\rho T_\mathrm{gas}}{\bar{m}} + \frac{1}{3}a_\mathrm{R} T_\mathrm{gas}^4 ~,
\end{equation}
where $\bar{m} = 0.615 m_H$ and $a_\mathrm{R} = 4\sigma/c$ is the radiation constant.  We can now solve for $T_\mathrm{gas}(r,z)$.  The initial azimuthal velocity is taken to be Keplerian, $V^\phi(r) = \Omega_\mathrm{K}$.  Note that we neglect additional corrections to the Novikov-Thorne solution\cite{Penna12}, but since we are just using these conditions to initialize our simulations, this should not matter too much.  For the background, we initialize a cold (gas internal energy density, $e = 10^{-4} e_\mathrm{max} r^{-5/2}$), low density ($\rho = 10^{-6} \rho_\mathrm{max} r^{-3/2}$) fluid with velocity $u^r = -(r_s / r)^{1/2}$ where $r_s$ is the Schwarzschild radius. The temperature found above is also used to set the radiation field. In the frame of the fluid, the initial radiation energy density is
\begin{equation}
E_\mathrm{rad} = a_R T_\mathrm{gas}^4 ~,
\end{equation}
while the flux, $F^i$, is set equal to the gradient of this quantity. Note that we assume the surface and burst radiation fields are isotropic. A limitation of the $\bf{M}_1$ closure is that the flux moves in a single direction. For the burst radiation, this is radially away from the star. Thus, we are neglecting effects related to scattering in the neutron star atmosphere, which would increase the intensity incident upon the disc by a factor of order $(1+2.06\mu)$ \cite{Chandrasekhar60,Lapidus85}, where $\mu$ is the cosine of the angle between the normal to the atmosphere and the direction of radiation. As a consequence, we are underestimating the effect of PR drag by a similar factor.

The simulations are 2.5 dimensional (azimuthal velocity and flux components are retained and evolved), axisymmetric, and cover a radial range from $r_\mathrm{min} = 10.7$ km to $r_\mathrm{max} = 352$ km, with exponential spacing. The assumption of axisymmetry is needed to keep the computational cost reasonable, though there are observations of bursts with modulations in their tails suggestive of non-axisymmetric effects\cite{Mahmoodifar16}. We use ``outflow'' boundaries at both the inner and outer radial boundaries, which effectively means we are assuming any boundary layer effects related to the neutron star surface happen inside $r_\mathrm{min}$. The full range of $\theta$ is covered with a latitude coordinate, $x_2$, related to $\theta$ by $\theta = x_2 + 0.45 \sin (2x_2)$, which concentrates resolution toward the midplane. Two layers of static mesh refinement further enhance the resolution in the region of interest. The base resolution is $96^2$, with the additional layers of refinement covering the region $r_\mathrm{min} \le r \le 64$ km and $0.35\pi \le x_2 \le 0.65\pi$ ($84^\circ \le \theta \le 96^\circ$) for an effective peak resolution of $384^2$ over the disc.

\end{methods}

\begin{addendum}
 \item[Data Availability]The source data required to reproduce all figures, except Fig. 5 and Fig. 10, are provided in the Source Data. The raw simulation data is available from the corresponding author upon reasonable request.
\end{addendum}

\begin{figure}
\begin{center}
\includegraphics[width=\textwidth]{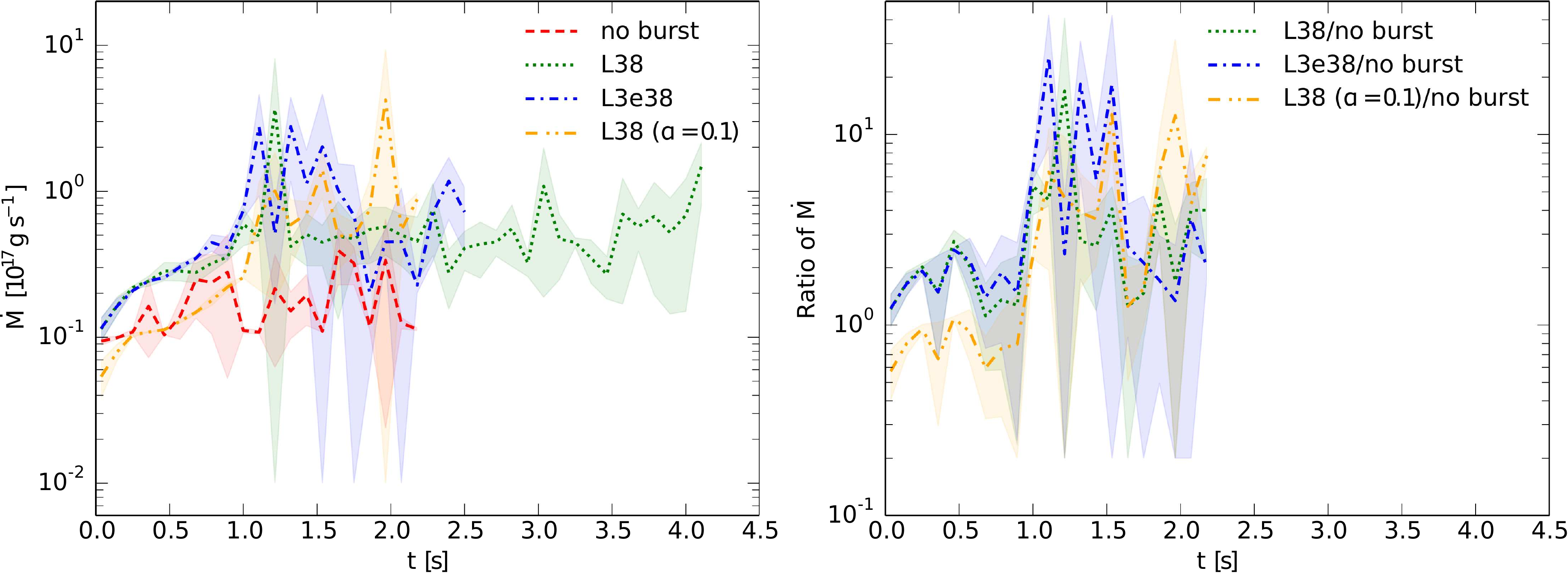}
\end{center}
\caption{{\bf $\dot{M}$ onto the neutron star.} {\it Left:} Moving average of the mass accretion rate, $\dot{M}$, measured near the inner radial boundary of the simulation domain for all four simulations. {\it Right:} Plot of the ratio of burst accretion rates to the non-burst rate. All burst simulations show sustained, enhanced accretion.}

\label{fig:mdot}
\end{figure}

\begin{figure}
\begin{center}
\includegraphics[width=\textwidth]{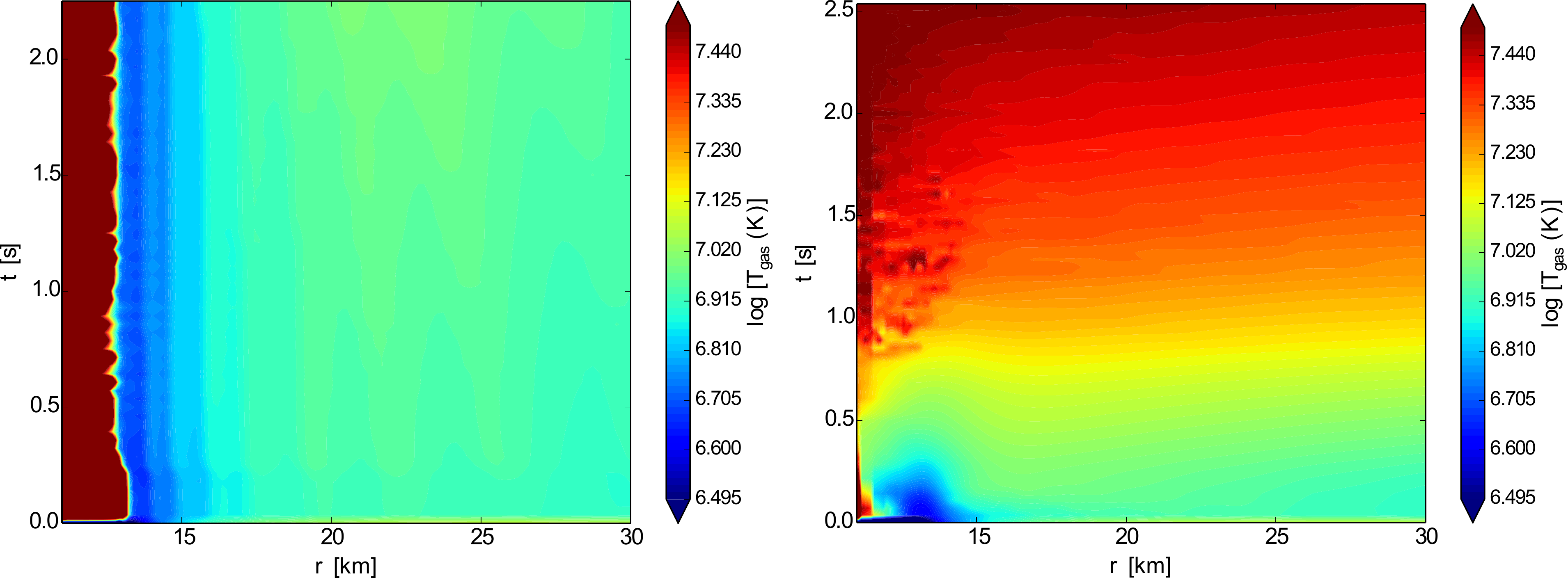}
\end{center}
\caption{{\bf Spacetime diagrams of the mass-weighted, shell-averaged gas temperature.} {\it Left:} from the simulation without a burst; and {\it Right:} from the simulation with a burst of $3 \times 10^{38}$ erg~s$^{-1}$. The burst causes significant heating over a wide radial range in the disc.\label{fig:T_gas}}
\end{figure}

\begin{figure}
\begin{center}
\includegraphics[width=\textwidth]{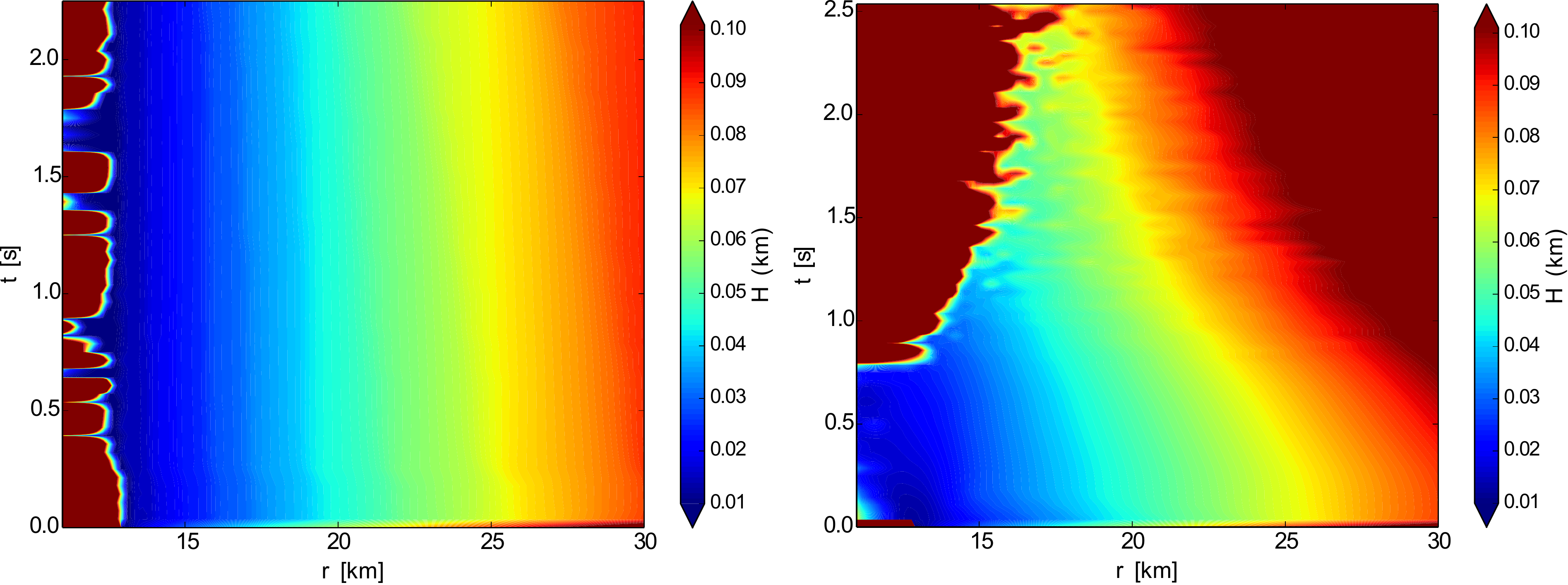}
\end{center}
\caption{{\bf Spacetime diagrams of the mass-weighted, shell-averaged disc height. } {\it Left:} from the simulation without a burst; and {\it Right:} from the simulation with a burst of $3 \times 10^{38}$ erg~s$^{-1}$. Sustained heating of the disc by the burst causes the disc height to increase over the timescale of the burst.\label{fig:height}}
\end{figure}

\begin{figure}
\begin{center}
\includegraphics[width=0.5\textwidth]{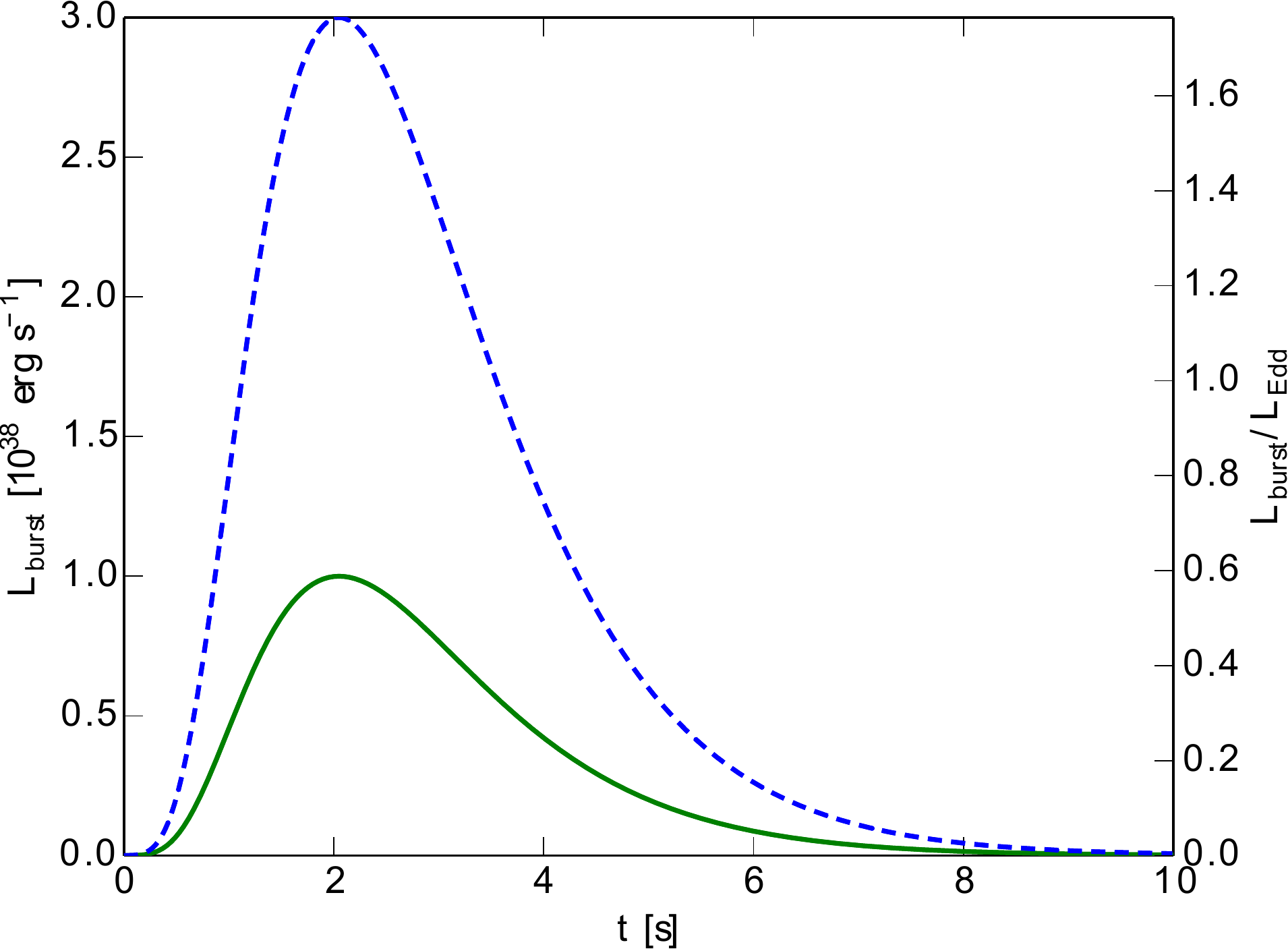}
\end{center}
\caption{{\bf Burst profiles.} Plots of the two different burst profiles used in this work, showing exponential rise followed by exponential decay.\label{fig:profiles}}

\end{figure}

\end{document}